# Enhanced Plasmonic Photocatalysis through Cooperative Plasmonic–Photonic Hybridization

Qinglan Huang,[1,2] Taylor D. Canady,[2,3] Rohit Gupta,[4] Nantao Li,[1,2] Srikanth Singamaneni,[4] and Brian T. Cunningham[1,2,3,5,*]

[1]Department of Electrical and Computer Engineering, University of Illinois at Urbana–Champaign, Urbana, IL 61801, United States;

[2]Holonyak Micro and Nanotechnology Laboratory, University of Illinois at Urbana–Champaign, Urbana, IL 61801, United States;

[3]Carl R. Woese Institute for Genomic Biology, University of Illinois at Urbana–Champaign, Urbana, IL 61801, United States;

[4]Department of Mechanical Engineering & Materials Science, Washington University in St. Louis, St. Louis, MO 63130, United States;

[5]Department of Bioengineering, University of Illinois at Urbana-Champaign, Urbana, IL 61801, United States

*Corresponding author: bcunning@illinois.edu

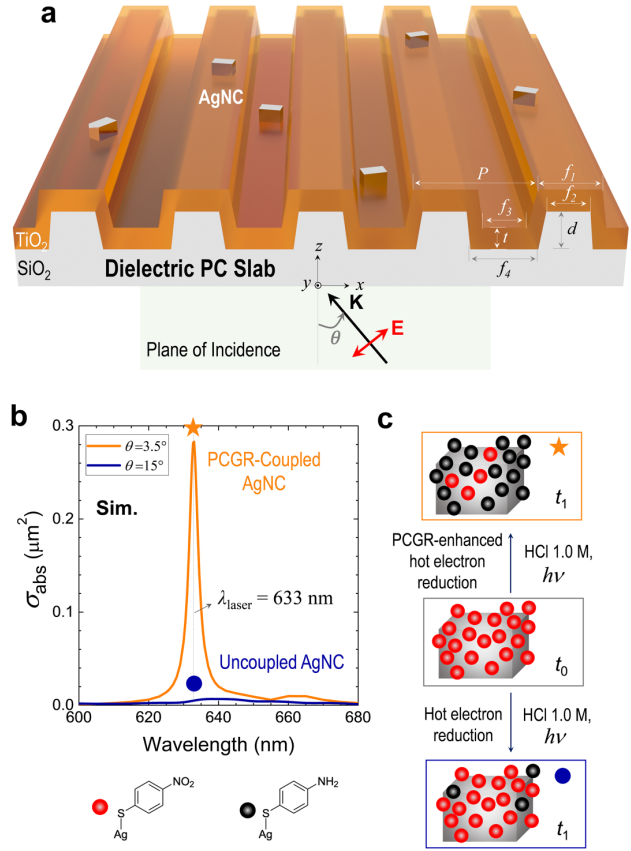

Fig. 1. Design of a nanophotonic chip for enhanced plasmonic photocatalysis. (a) Plasmonic photocatalyst silver nanocuboids (AgNCs) are integrated onto an on-resonant dielectric photonic crystal (PC) slab to form a plasmonic–photonic hybrid. A 3-nm thick $SiO_2$ isolation film covering the PC surface is omitted in the schematic. The excitation laser ($\lambda_{laser}$ = 633 nm) is TM-polarized (in-plane **E** field) with an incidence angle $\theta$. Structure parameters of the PC slab: $P$ = 380 nm, $d$ = 130 nm, $t$ = 76.7 nm, $f_1$ = 0.521, $f_2$ = 0.356, $f_3$ = 0.45, $f_4$ = 0.61. (b) Simulated absorption cross section $\sigma_{abs}$ of an individual AgNC on the PC surface, when it is coupled ($\theta = 3.5°$, orange) or uncoupled ($\theta = 15°$, navy) to the PC guided resonance (PCGR). (c) In the presence of 1.0 M HCl, the 4-nitrothiolphenol (4-NTP) chemisorbed on AgNCs undergo a hot-electron-mediated reduction to form 4-aminothiophenol (ATP). The PCGR-coupled conversion rate ($\theta = 3.5°$, orange box) is much higher than that of the uncoupled cases ($\theta = 15°$, navy box). Bottom left corner: colored labels for 4-NTP and 4-ATP.

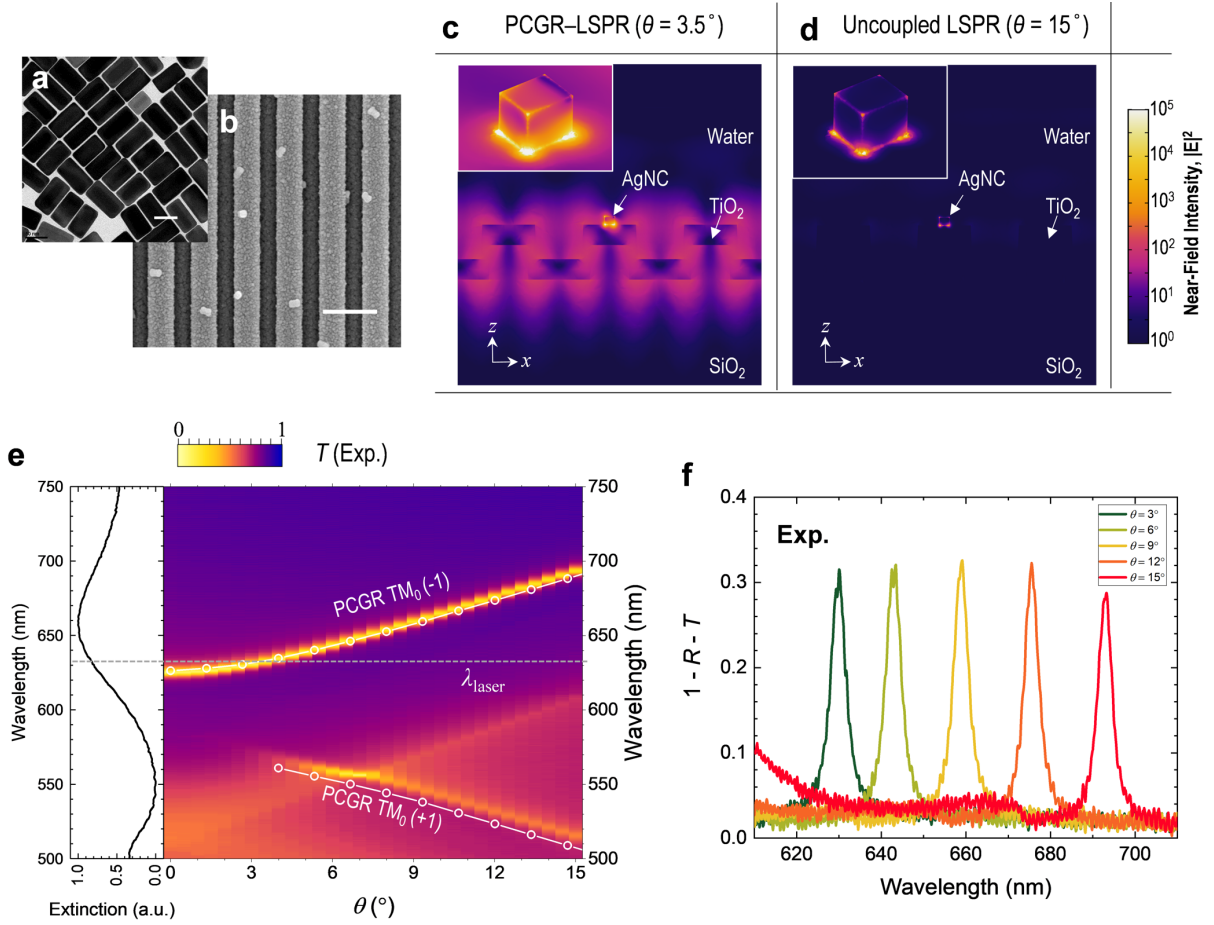

Fig. 2. Optical properties of the LSPR–PCGR hybrid mode. (a) TEM image of the AgNC. Scale bar 50 nm. (b) Representative SEM image of the AgNC–PC hybrid structure. Scale bar 500 nm. (c-d) Simulated near-field intensity ($|\mathbf{E}|^2$, normalized to incidence field) of the AgNC–PC hybrid at (c) $\theta = 3.5°$ and (d) $\theta = 15°$. The $xz$ cross-sectional slice cuts through the front surface of the AgNC. The insets show a close-up 3D view of the AgNC surface. (e) Left panel: measured extinction spectrum of AgNCs on a TiO$_2$-coated glass substrate immersed in water. Right panel: angle-resolved transmission spectra of the AgNC–PC hybrid. The simulated spectral positions of the two counterpropagating PCGRs are overlaid (white line with circular symbols). The horizontal dashed line indicates the spectral position of the excitation laser. (f) Measured (1-reflectance-transmittance) efficiency spectra of the AgNC–PC hybrid at various incidence angles.

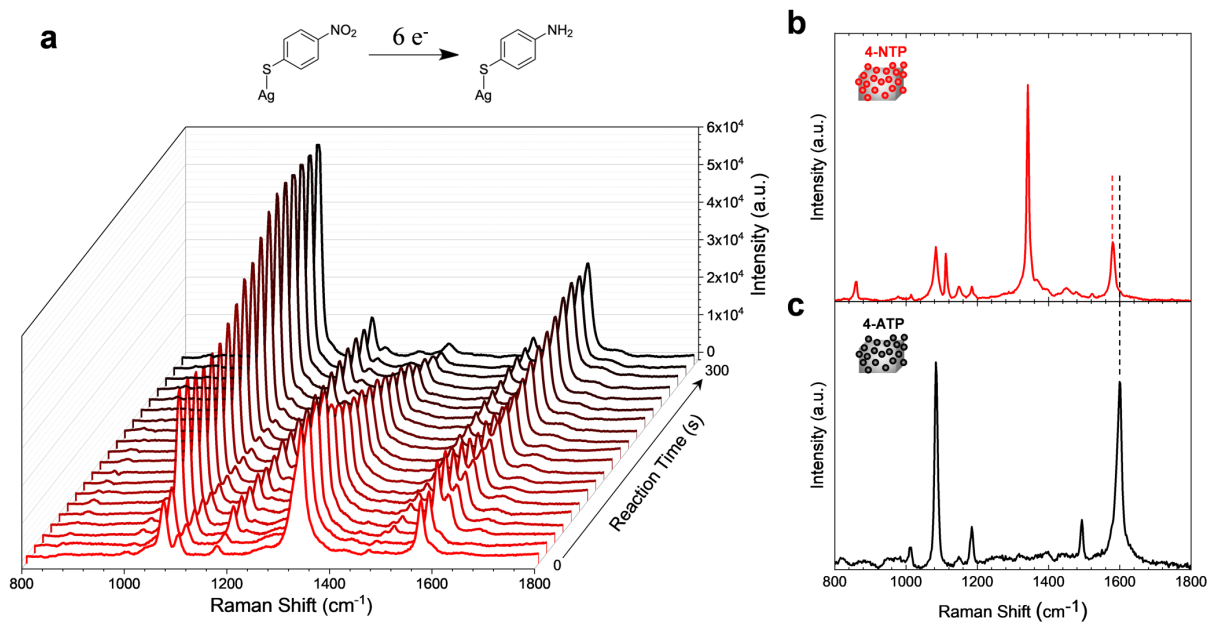

Fig. 3. Real-time observation of the hot-electron-driven reaction via SERS. (a) Time-dependent SERS spectra of 4-NTP-modified AgNCs placed on a PC slab in the presence of 1.0 M HCl. Excitation parameters: incidence angle $\theta = 3.5°$, laser power 5 mW, illumination area 0.5 mm × 5 $\mu$m, integration time 1 s. The gradual line color change from red to black represents the progressive transition of 4-NTP to 4-ATP. Reference SERS spectra of (b) 4-NTP and (c) 4-ATP obtained in control experiments. The scales of vertical axes for (b) and (c) are not the same. The fingerprint Raman bands used to estimate conversion efficiency are highlighted.

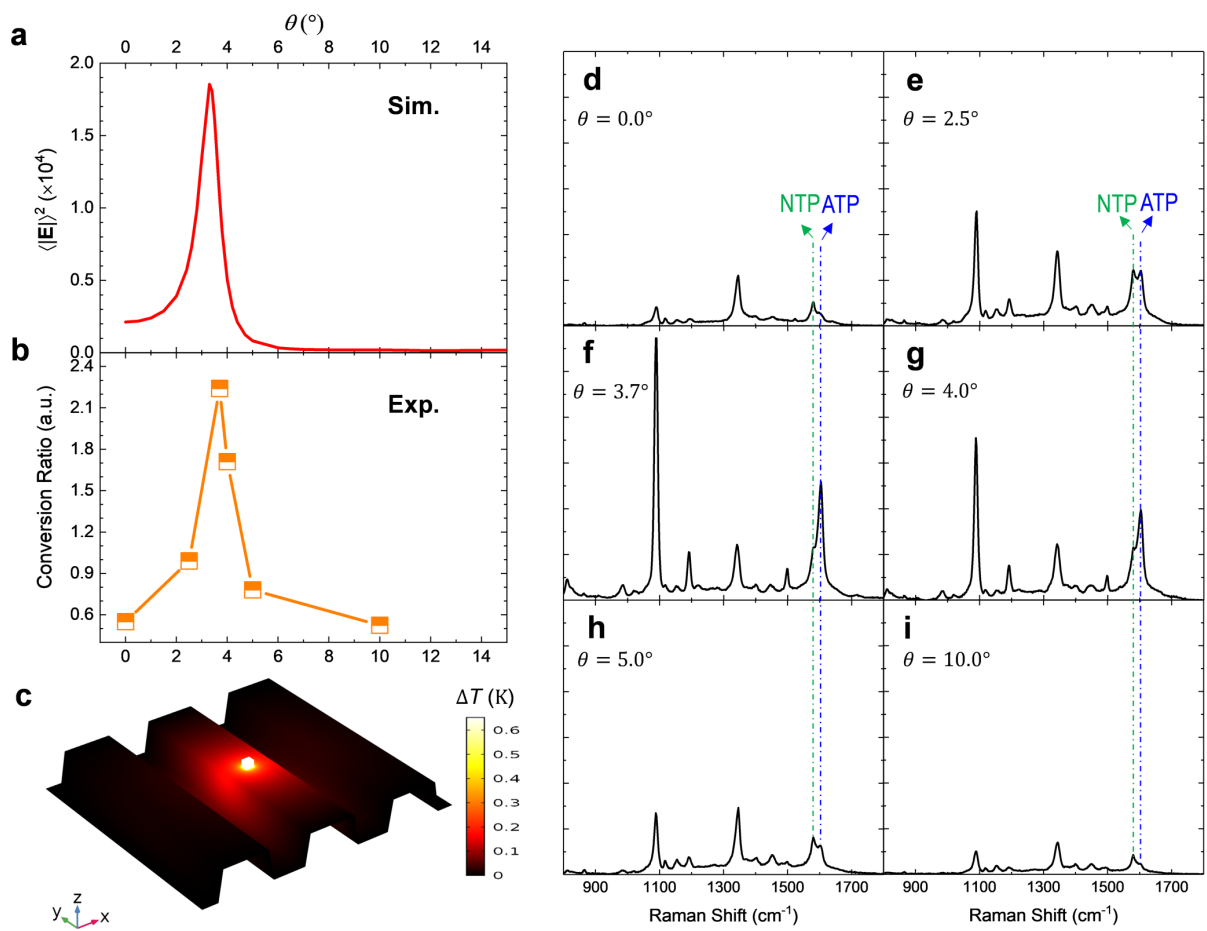

Fig. 4. Enhanced hot electron photocatalysis in plasmonic–photonic coupling. (a) Simulated average near-field intensity on the surface of the AgNC (<|**E**|$^2$>, normalized to incidence field) for $\lambda_{laser}$ = 633 nm as a function of incidence angle $\theta$. (b) Experimentally obtained reaction conversion ratio (defined by the ratio of SERS intensity at 1599 cm$^{-1}$ and 1580 cm$^{-1}$) after 180 s illumination as a function of $\theta$. (c) Simulated temperature (relative to the room temperature 293.15 K) distribution of the proposed photocatalyst surface at the LSPR–PCGR hybridization. (d-i) The SERS spectra after reaction at each $\theta$ denoted in (b). The Raman peaks utilized for calculating the conversion ratio are highlighted.

**Abstract**

Plasmonic nanoparticles (NPs) hold tremendous promise for catalyzing light-driven chemical reactions. The conventionally assumed detrimental absorption loss from plasmon damping can now be harvested to drive chemical transformations of the NP adsorbent, through the excitation and transfer of energetic "hot" carriers. The rate and selectivity of plasmonic photocatalysis are dependent on the characteristics of the incident light. By engineering the strength and wavelength of the light harvesting of a NP, it is possible to achieve more efficient and predictive photocatalysts. We report a plasmonic–photonic resonance hybridization strategy to substantially enhance hot electron generation at tunable, narrow-band wavelengths. By coupling the plasmon resonance of silver NPs to the guided mode resonance in a photonic crystal (PC) slab, the reaction rate of a hot-electron-driven reduction conversion is greatly accelerated. The mechanism is broadly compatible with NPs with manifold materials and shapes optimized for the targeted chemistry. The novel enhancement platform sheds light on rational design of high-performance plasmonic photocatalysts.

**Introduction**

A new paradigm of plasmonic photocatalysis has emerged as a platform for triggering energetically intensive chemical reactions under mild temperature conditions and with potentially selective reaction pathways control.[1] Plasmonic nanoparticles (NPs) have been shown to drive molecular desorption, bond cleavage, and single- and multi-electron redox reactions on their surface.[2] Through the excitation of localized surface plasmon resonance (LSPR), a metal NP doubly functions as a nanoantenna that confines optical energy into sub-diffraction volumes,[3] and as a reactive element that interacts with the adsorbed molecules.[4] As a mixed light–matter mode, a plasmon polariton (PP) partially stores its energy in the kinetic motion of the free carriers, which leads to inevitable dissipative loss.[5] PPs can decay through chemical interface damping where they directly excite a carrier from the metal to the molecule.[6] Alternatively, they can decay by exciting electron–hole pairs in the metal through several different mechanisms (namely, interband absorption, phonon and defect assisted absorption, electron–electron scattering assisted absorption, and Landau damping (or surface collision assisted absorption)).[5] Ultimately, for productive chemistry to occur, the excited energetic carriers must be transferred to the adsorbed molecules before ultrafast carrier relaxation processes.[4] In this way, the energy in the LSPR can be deposited into an adsorbed molecule, driving it to a new, excited potential energy surface (PES) of a chemical reaction.[7]

One central quest in this emerging field is to enhance the reactivity of a plasmonic photocatalyst. The plasmonic catalytic activity is a convolution of many effects, including plasmon-derived phenomena such as near-field enhancement, absorption, and heating. In addition, the NPs' catalytic activity in the dark and lattice-derived mechanical vibrations (phonons) must be accounted for.[7] Independent engineering of those properties of a NP would allow additional degrees of freedom in the design of efficient plasmonic photocatalysts. As the energy quanta required to initiate a chemical reaction are supplied by plasmon decay, an obvious strategy is to enhance the NP's absorption efficiency at the excitation wavelengths.

Manipulating the NPs' absorption characteristics would have deep impacts on the reaction kinetics. The supralinear illumination intensity dependence of photocatalytic reaction rate in multiple plasmon-driven reactions indicates higher quantum yield with increasing photon flux.[8,9] Specifically, the activation barrier of plasmonic photocatalysis has been shown to be a function of excitation wavelength and intensity.[10,11] This is because more energetic carriers are transferred into or directly excited into the metal–molecule surface species at an increased photon flux at the LSPR wavelength, which brings the adsorbate to an excited electronic state of the PES.[12] The excited state of the PES can have a decreased activation barrier associated with increased reaction rate and efficiency,[10] or new valleys that offer selective reaction pathways not accessible in thermally-driven catalysis.[13] Moreover, multi-carrier photoredox reactions, which are central to artificial photosynthesis but are kinetically sluggish,[14] are only possible under very intense laser excitation, because two or more hot electrons must be simultaneously generated in a NP.[15] Intriguingly, amplifying the absorption cross section of the NPs is equivalent to an increase in the photon flux, so the capability to manipulate mechanistic details of the reactions follows without increasing the illumination power.

Whereas the reactivity of a plasmonic NP can be optimized through material choice (for example using multiple metals which separately function as the antenna and reactor sites),[9,10,16] size and shape (small radius of curvature is beneficial for charge transportation),[17,18] the hot carrier generation (absorption) can be independently modified through engineering its photonic environment.[19-21] Here we show that the plasmon-derived hot carrier generation can be significantly enhanced when hybridized with a photonic microcavity resonance. The kinetics of hot-electron-driven redox reactions on silver nanocuboids (AgNCs) are greatly improved when they are evanescently coupled to a photonic crystal guided resonance (PCGR),[22] compared to the uncoupled AgNCs control, and the reaction rate is directly dependent on the angle-resolved LSPR–PCGR near-field enhancements.

**Results**

We introduce a versatile platform capable of enhancing the hot carrier generation in plasmonic photocatalysts at a tunable, narrow spectral band. The structure (Fig 1(a)) is composed of AgNCs randomly adsorbed on the surface of a dielectric photonic crystal (PC) slab.[23] The PC slab supports counter-propagating PCGRs in a $TiO_2$ thin film ($n = 2.35$) coated on a periodically modulated glass substrate ($n = 1.47$). The surface is immersed in aqueous media, and the backside is excited with a transverse magnetic- (TM-) polarized laser ($\lambda_{laser} = 633$ nm) at incidence angle $\theta$. The spectrally overlapped LSPR and PCGR can resonantly couple via evanescent fields to form a synergistic plasmonic–photonic hybrid mode.[24-27] The angle-specific mode hybridization impacts the plasmon-driven reaction by delivering a sharp absorption enhancement at the LSPR. Our simulations predict that PCGR-coupling can amplify the absorption cross section of the AgNC by ~35× with a line width of ~4 nm (Fig. 1(b)). The total number of hot carriers excited in the process of plasmon absorption would hence be increased in the LSPR–PCGR hybridization. We will quantify this effect by observing a hot-carrier-driven redox reaction via surface enhanced Raman scattering (SERS), in which 4-nitrothiophenol (4-NTP) molecules chemisorbed on AgNCs undergo a six-electron-mediated reduction to form 4-aminothiophenol (4-ATP) in the presence of HCl.[17, 28] The conversion rate is expected to be angle-dependent, exhibiting higher catalytic activity for PCGR-coupled AgNCs, compared to uncoupled cases (Fig. 1(c)).

We synthesized AgNCs comprised of gold nanorod (AuNR) as cores and silver as shells,[29] as shown in the transmission electron microscopy (TEM) image in Fig. 2(a). These NPs are chosen as the plasmonic photocatalysts in this study because of their (1) strong LSPR in the red spectral range compatible with our laser and PC slabs (Fig. S1), (2) sharp edges and tips that facilitate hot carrier transportation,[17, 30] and (3) reactive silver material specified for the reaction under study. The AgNCs are surface-functionalized with 4-NTP molecules and uniformly deposited onto a PC surface without aggregation (Fig. 2(b)). A sparse coating of 1-3 AgNCs/$\mu m^2$ has been shown to lead to optimized hybrid enhancements, where critical coupling between the LSPR and PCGR leads to maximum energy absorption in NPs.[25, 31] The AgNCs are randomly orientated, and those aligned along the $x$-axis (matching the excitation field polarization) would be optimally activated.[24] To avoid losing hot electrons to the underlying $TiO_2$ which is commonly used as an "electron filter",[32, 33] a 3-nm-thick $SiO_2$ layer was sputtered onto the PC surface before AgNC deposition to electronically isolate them from the $TiO_2$.

The hybrid mode is activated when satisfying the phase matching condition of the PCGR mode, whereas the LSPR mode is generally angle insensitive. This incidence angle selection rule allows direct comparison of the absorption, near-field intensity, SERS intensity, and catalytic activity between the hybrid supermode and the solitary LSPR mode. A near-field picture is helpful for understanding the coupling behavior. The hybridization ($\theta = 3.5°$) is characterized by a standing wave pattern in the PC slab and an intense optical hotspot concentrated on the AgNC (Fig. 2(c)). The PCGR-coupled AgNC possesses a strongly enhanced electric field compared to the uncoupled case at a detuned angle (Fig 2(d)). In contrast to the plasmonic gap modes[34] where the hotspots are only accessible to a small region of the bridged NPs, our open cavity offers amplification across all the reaction sites over the NP surface. In essence, the coupling acts as an impedance matching network that cooperatively combines the cavity's quality factor and the antenna's mode confinement.[25, 31] Energy of incident photons from free space successively oscillates in a photonic microcavity, concentrates into PPs, decays into hot carriers and finally transfers into molecules to alter their chemical composition. The intense electric field confined at the AgNC is the physical phenomenon allowing for Landau damping, which has been identified as the key mechanism to generate highly energetic carriers right at the surface, where they have access to the analyte.[35]

Moreover, the spectrally tunable narrowband hybrid resonance allows matching to specific electronic transitions and production of hot carriers with defined energy. Each wavelength corresponds to a distinctive mechanism of PP decay, leading to generation of hot carriers with different properties and relaxation cascade, and ultimately transferring different energy to the molecules.[36] Wavelength is therefore a key factor influencing the reaction rate, selectivity, and pathways,[10, 14] and the spectral tunability of the hybrid resonance offers additional accessibility in selective reactions. Derived from the band diagram of the PCGRs, the sharp resonance absorption can be tuned over a broad wavelength range to cover different absorption mechanisms by scanning the incidence angle (Fig. 2(e) right panel). In this work we target at Landau damping at the LSPR wavelength (Fig. 2(e) left panel) with $\lambda_{laser}$ = 633 nm. The predicted enhanced absorption effect is validated by measuring the angle-resolved extinction spectra 1-$R$-$T$, where $R$ and $T$ are the zeroth-order reflection and transmission efficiencies, respectively. Narrowband extinction peaks at the PCGR-prescribed wavelengths are ~20× larger than that of the uncoupled LSPR, which can be obtained when the incidence angle is detuned (Fig. 2(f)). Small deviations from the simulation (Fig. 1(b)) can be attributed to their discrepancy in AgNC density and orientations.

Once the enhanced optical activity was confirmed, we performed hot-electron-driven surface chemistry as described in Fig 1(c). As a thiolated molecule, 4-NTP forms a densely packed monolayer on the AgNC surface.[37] The LSPR excited in the AgNC undergoes nonradiative plasmon decay to generate hot carriers. By transferring six energetic electrons to a chemisorbed 4-NTP, its nitro terminal group can be on-site reduced to an amino group.[17] An acid halide media (that is, HCl, HBr, and HI) is necessary to trigger this redox reaction, in which protons act as the hydrogen source and halide anions act as a hole scavenger. The counter-half reaction involves the formation and subsequent photodissociation of insoluble silver halide to regenerate the silver surface.[28] Additionally, a series of control experiments have demonstrated that the reaction is induced neither by photothermal by nor photochemical effects.[17, 28]

We observe the molecular conversion process through real-time SERS monitoring.[24] A homebuilt inverted line-focusing Raman microscope[38] is used to stimulate the reaction on the AgNC–PC hybrid, and concurrently collect Raman scattered photons during the process. Briefly, the incident beam is focused into a line (~5 $\mu$m × 0.5 mm) oriented along the *x*-axis on the sample surface, and the incidence angle can be precisely adjusted. This arrangement allows for the efficient excitation of the PCGR–LSPR hybrid mode, as well as simultaneous excitation of multiple AgNCs in an area considerably larger than a point focused spot.

The PC slab decorated by 4-NTP-modified AgNCs is immersed in 1.0 M HCl and excited at the resonance angle ($\theta = 3.5°$). The laser was fixed at the same spot to investigate the molecular evolution over time, and the successive SERS spectra are depicted in Fig. 3(a). To establish a reference, we also measured the SERS spectra of the 4-NTP- and 4-ATP-functionalized AgNCs placed on PC slabs, but without the acid media as shown in Fig. 3(b) and (c), respectively. As expected, no reaction was observed over time in the absence of acid halide. The time-dependent

SERS spectra of the reaction exhibited characteristic bands at 1083, 1341, and 1580 cm$^{-1}$ at the beginning of the transformation, which were respectively assigned to C–S stretching, O–N–O stretching, and the phenyl-ring mode of 4-NTP.[37] The intensities of the R-NO$_2$-associated bands progressively decreased with the conversion of 4-NTP to 4-ATP, and concomitantly two characteristic bands of 4-ATP at 1493 and 1599 cm$^{-1}$ emerged.[39] After 5 min illumination, most of the 4-NTP molecules chemisorbed on the AgNCs are converted into 4-ATP.

We then investigated the effect of plasmonic–photonic hybridization on plasmon-assisted catalytic activity by driving the reaction at various incidence angles (but the same laser power). By properly selecting the illumination time, a divergence in the reaction rates is expected for different angles. We prepared multiple small polydimethylsiloxane (PDMS) wells that contain 1.0 M HCl media on a hybrid photocatalyst surface, and we illuminated each well at a distinct $\theta$ for the same amount of time. Since AgNCs are uniformly distributed across the 1 × 1 cm$^2$ PC slab surface, it is fair to compare the reaction rates among different wells. The SERS spectra after 180 s illumination under each $\theta$ are measured and shown in Fig. 4 (d-i). The difference in overall SERS intensity when scanning $\theta$ is a direct result of near-field enhancement via plasmonic–photonic hybridization. Fig. 4(a) shows the simulated evolution of the average electric field intensity ($\langle |\mathbf{E}|^2 \rangle = \iint |\mathbf{E}|^2 \, dS / \iint dS$) at $\lambda = 633$ nm as a function of incidence angle. Formally, the SERS enhancement is proportional to the product of electric field intensities at the excitation frequency and the Raman scattering frequency.[40] In our system the SERS intensity scales only with the first term because the photonic environment for the scattered photons remains unmodified when changing excitation angles,[24] and this can be clearly seen with the angle-resolved SERS intensity of 4-NTP-modified hybrid structure in absence of acid halide (Fig. S4).

The relative intensity of the SERS bands of 4-NTP and 4-ATP reflects their surface coverage rate, and we can use the ratio of 1599 and 1580 cm$^{-1}$ band intensity to "quantify" the conversion efficiency. We note that the intensity ratio between their SERS signals does not directly translate into molecule counts because 1) the Raman cross-section ($\sigma$) for 4-NTP and 4-ATP are not the same ($\sigma_{\mathrm{NTP}} > \sigma_{\mathrm{ATP}}$),[17] and 2) the result is masked by the near-field distribution (SERS enhancement factor) across the AgNC surface. As seen in Fig. 4(d-i), there are remarkable discrepancies in the apparent conversion efficiencies under different illumination angles: higher rate when activating the LSPR–PCGR hybrid resonance at $\theta = 3.7°$, compared to LSPR operating alone when $\theta$ is detuned. Specifically, Fig. 4(b) depicts the change in the apparent conversion rate as a function of $\theta$. Not surprisingly, we find a close correlation between the near-field enhancements (Fig. 4(a)), SERS enhancements (Fig. S4), and plasmonic photocatalysis rate (Fig. 4(b)). This result provides evidence that the increased plasmon-assisted catalytic activity stems from the electromagnetic enhancements through LSPR–PCGR hybridization.

While hot electron excitation can be synergistic with the thermo-plasmonic effects in plasmonic photocatalysis, in this study the hybridization-induced catalysis activity enhancement has an electronic origin. We conducted thermal simulations (COMSOL) to estimate the temperature in the AgNC upon its hybridization with the PC slab, when the heat generation is the strongest. Briefly, the power adsorbed by the AgNC (obtained by the product of absorption cross section and illumination intensity) was input into a steady-state heat transfer model as a volumetric heat source.[41, 42] The temperature profile is shown in Fig. 4(c). Due to the good thermal conductivity of the environment and the low power of the CW laser (5 mW), only a small temperature rise (~0.65°C) is induced in the AgNC. This mild thermal effect is unlikely to have any significant effects on the chemical reaction. We therefore ascribe the accelerated reaction rate to the increased hot electron generation rate through the cooperative plasmonic–photonic coupling. The increased rate of energetic carriers transferred to 4-NTP speeds up the reduction process.

**Discussion**

In summary, we have demonstrated a new mechanism to enhance the plasmon-assisted catalytic activity through coupling plasmonic NPs to a photonic microcavity. Through forming an intense optical hotspot at the NP, the LSPR–PCGR hybridization significantly enhances the hot carrier generation at selected narrowband wavelengths with broad spectral tunability. This platform will pave the way to high-performance plasmonic photocatalysis because (1) it is widely compatible with a variety of NPs, allowing its optical and chemical properties to be optimized individually for targeted reactions; (2) its enhanced light harvesting represents a critical step toward energy-efficient photocatalysts which can be driven by low power lasers, light emitting diodes, or even sunlight; (3) its high hot-carrier generation rate sheds light on the kinetically challenging multi-carrier reactions and holds promise for activation barrier reduction; and (4) its narrow, spectrally-tunable absorption enhancements allow for discrete excitement of defined electronic transitions in the metal, which offers a degree of freedom to manipulate the selectivity of the reactions.

**Acknowledgements**

This work was supported by National Science Foundation grant NSF CBET 19-00277. Q.H. acknowledges the Sah Doctoral Fellowship from UIUC ECE. T.D.C. is supported by an Institute for Genomic Biology (IGB) fellowship.

**Supporting Information**

Materials and methods (fabrication of PC slabs, synthesis of AgNCs, AgNC modification and characterization, integration of AgNCs with PC slabs, far-field characterizations), numerical simulations (electromagnetic and thermal modeling using COMSOL), and measurement of angle-resolved SERS (optical setup, SERS as a function of incidence angles) are detailed in the supporting information.